\documentclass[final,5p,times,twocolumn]{elsarticle}
 \usepackage{epsfig}

\usepackage{amsfonts}
\usepackage{mathrsfs}
\usepackage{bm}

\biboptions{sort&compress}

\usepackage[usenames,dvipsnames]{xcolor}

\definecolor{R}{rgb}{0.9,0.0,0.1}
\definecolor{G}{rgb}{0.3,0.7,0.345}
\definecolor{Y}{RGB}{255,255,0}

\makeatletter
\def\ps@pprintTitle{%
  \let\@oddhead\@empty
  \let\@evenhead\@empty
  \def\@oddfoot{\reset@font\hfil\thepage\hfil}
  \let\@evenfoot\@oddfoot
}
\makeatother

\def\be{\begin{equation}}
\def\ee{\end{equation}}
\def\bea{\begin{eqnarray}}
\def\eea{\end{eqnarray}}
\def\bfl{\begin{flushleft}}
\def\efl{\end{flushleft}}
\def\bfr{\begin{flushright}}
\def\efr{\end{flushright}}
\def\bc{\begin{center}}
\def\ec{\end{center}}
\def\ben{\begin{enumerate}}
\def\een{\end{enumerate}}
\def\bit{\begin{itemize}}
\def\eit{\end{itemize}}

\def\dzn{,\kern-0.1em,}
\def\lan{\langle \langle}
\def\ran{\rangle \rangle}

\newcommand{\tm}[1]{\mathrm{#1}}

\def\lan{\langle \langle}
\def\ran{\rangle \rangle}

\journal{Solid State Communications}

\begin{document}

\begin{frontmatter}

 \title{Neel temperature\tnoteref{label1}}

\title{ Phase diagram of spin-$\frac 12$ quantum Heisenberg  $J_1-J_2$ antiferromagnet
on the body-centered-cubic lattice in random phase approximation}

\author{Milan R. Panti\' c  \corref{cor}}
\ead{mpantic@df.uns.ac.rs} 

\author{Darko V. Kapor, Slobodan M. Rado\v sevi\' c, Petar M. Mali}

\cortext[cor]{Corresponding author}

\address{Department of Physics, Faculty of Sciences, University of Novi Sad, 
Trg Dositeja Obradovi\' ca 4, Novi Sad, Serbia}

\begin{abstract}

Magnetic properties of spin $\frac 12$ $J_1-J_2$ Heisenberg antiferromagnet on body centered cubic lattice
are investigated. By using two-time temperature Green's functions, sublattice magnetization
and critical temperature depending on the frustration ratio $p=J_2/J_1$ are obtained in both stripe and N\'eel
phase. The analysis of ground state sublattice magnetization and phase diagram
indicates the critical end point at $J_2/J_1 =0.714$, in agreement with previous studies.
\end{abstract}

\end{frontmatter}

\section{Introduction}
\label{intro}

 We are witnessing the increased interest
 in the study of  quantum phase transitions \cite{Sachdev,Wen}
 in magnetic systems \cite{Diep}, which can be triggered by varying  some  
of the system's parameters. These include  exchange integrals, external magnetic field, or the
doping concentration in the case of  high- Tc superconductors. 

The  motivation  for  studying  the quantum phase transitions comes in part from closely 
related problem  of strongly correlated systems at low temperatures. 
They display transition between ground state ($T=0$ K) phases  as a combination 
of the quantum fluctuations and the competition between interactions, i.e. 
frustration \cite{FM}. 
Nowdays, a frustrated spin systems are among the most interesting
and challenging topics in theoretical magnetism, including  even frustrated 2D Ising
model  \cite{Sandvik}. Competition between
exchange interactions in magnetic materials can lead to a 
variety of the magnetic ordering states, and  even to induce a phase
transitions between them. Consequently, the frustrated quantum Heisenberg magnets
with competing nearest - neighbour (NN) and next - nearest -
neighbour (NNN) antriferromagnetic (AF) exchange interactions
($J_1$ and $J_2$, respectively) have become extremely active field
of research.

The early works focused on the square lattice Heisenberg model, and  it was investigated in detail
by different methods \cite{2,2-1,3,4,5,6,7,8,9,10,11,NoviPRB,ReadSachdev}.
With vanishing  NNN interaction ($J_2= 0$),  the ground is
state known to be  antiferromagnetic (AF) for $S\geq 1$ \cite{Manousakis}. 
A nonzero NNN  interaction
 leads to the frustration and the ''crash'' of simple  AF ordering.
For large enough $J_2$, the "stripe phase" emerges \cite{ReadSachdev},
with possible spin-liquid phase at $J_2/J_1 \approx 0.5$.
(See \cite{Fradkin} and references therein).  We shall
be interested in 3D version of the problem, 
with localized  spins
arranged  on a
body-centered-cubic (bcc) lattice.
The mean field calculation \cite{Smart} indicates existence of  two phases, represented
with different spin orderings on the Figure 1.  N\' eel state (AF1) can be described by a
standard two-sublattice system, while the proper treatment of 
the stripe phase (AF2)
can be accomplished  by the introduction of four
sublattices.

\begin{figure*}
\bc 
\includegraphics[scale=0.75]{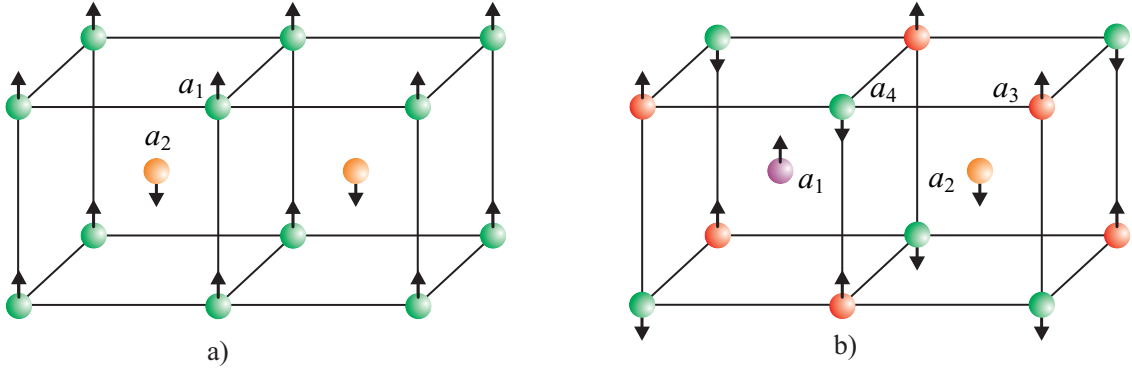}
\caption
{\label{Fig_1} (Color online) 
a) N\' eel ordering denoted as AF1 phase in further text. b) stripe ordering denoted as AF2 phase in further text
(after \cite{Smart}).}
\ec
\end{figure*}

The transition between two phases is governed by the frustration ratio
$p=J_2/J_1$ and  the mean-field
calculation gives  $p_{\tm{MF}} = 2/3$  for its 
critical value \cite{Oitma}. Numerous sophisticated methods
were subsequently used to obtain more reliable value 
of the critical frustration ratio.
Schmidt et al.
\cite{Schmidt} carried out exact diagonalization of finite 3D
lattices with periodic boundary conditions. Ground
state energy was found to have a discontinuity at $p = 0.693$ indicating a first
order quantum transition between two phases. This was confirmed by
calculation of magnetization and extrapolation to infinite
lattice. J. Oitmaa and W. Zhang \cite{Oitma} performed high-order
linked-cluster expansion at $T = 0 K$ and obtained that two
branches of ground state energy for AF1 and AF2 phases cross at $J_2/J_1
= 0.705 \pm 0.005$.
Majumdar and Datta \cite{Majumdar1} presented a non-linear spin
wave theory (up to quartic terms in Bose-operators) giving quite
 similar results. Majumdar \cite{Majumdar2} also extended this problem
to the antiferromanget on stacked square lattices with different
exchange in vertical direction.

The aim of this paper is to present, up to the our knowledge, the first application of
spin operator Green's functions on spin-$\frac 12$ bcc lattice  $J_1-J_2$ Heisenberg
antiferromagnet. It will be shown that spin operator temperature Green function
(TGF) method, in combination with random phase approximation (RPA),
yields  reliable  results both at $T=0$K and at critical temperature.
The paper is organized as follows. 
The  RPA magnon spectrum in both AF1 and AF2 phase
is determined
 in the Section \ref{MagnSpectr}. The calculation of ground state sublattice magnetization
is presented  in   Section \ref{SublMagn}, while Section \ref{NeelPhaseDiag}
contains discussion on the N\' eel temperature and phase diagram.
Finally, Section \ref{Concl} concludes the paper.

\section{Magnon Spectrum} \label{MagnSpectr}

\subsection{AF1 (N\' eel) phase}

We start with the spin-$\frac 12$ Hamiltonian of the system with two sublattices
$a_1$ and  $a_2$  (See Figure \ref{Fig_1} a)):
 \bea
\hspace*{-0.8cm}\hat{H}_{\tm{AF1}} & = & J_1 \sum_{\bm n, \bm \lambda} 
 \hat{{\color{G}\bm S}}_{\bm n}^{(a_1)} \cdot  \hat{{\color{BurntOrange}\bm S}}_{\bm n +\bm \lambda}^{(a_2)}
+ \frac{J_2}{2} \sum_{\bm n, \bm \delta} 
 \hat{{\color{G}\bm S}}_{\bm n}^{(a_1)} \cdot  \hat{{\color{G}\bm S}}_{\bm n +\bm \delta}^{(a_1)} 
 \nonumber \\
& + & \frac{J_2}{2} \sum_{\bm n, \bm \delta} 
 \hat{{\color{BurntOrange}\bm S}}_{\bm n}^{(a_2)} \cdot  \hat{{\color{BurntOrange}\bm S}}_{\bm n +\bm \delta}^{(a_2)} 
 \label{2.5}.
  \eea
Here $\bm n$ denotes the site in the given sublattice, each having $N_{\tm{N}}=N/2$ sites
and both exchange parameters $(J_1, J_2)$ are assumed to
be positive.
$\bm \lambda$ connects first neighbours from sublattices $a_1$ and $a_2$,
while $\bm \delta$ relates corresponding second neighbours.
It is seen from Figure \ref{Fig_1} that every site has $z_1 = 8$ nearest neighbours 
and that the number of second neighbours  is $z_2 = 6$.
To simplify calculations
we rotate operators from   $a_2$ sublattice about $S^x$ axis by
 $\pi$ as in \cite{EPJB,SSCTN}.

\begin{figure}
\bc 
\includegraphics[scale=0.55]{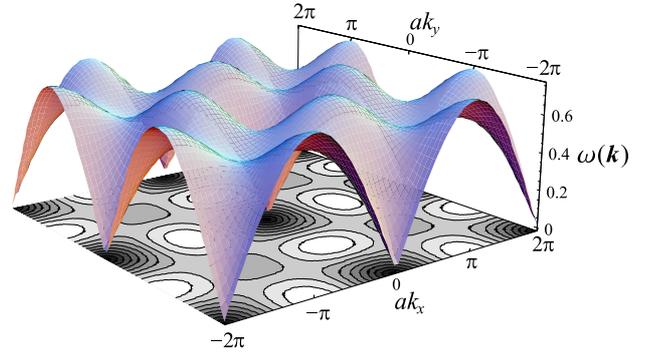}
\caption
{\label{NEl3D} (Color online) 
Reduced magnon energies $\omega (\bm k) = E(\bm k)/(J_1 z_1 \langle S^z \rangle)$ 
for $k_z = 0$, $p = 0.5$ in AF1 phase.}
\ec
\end{figure}

Writing down four equations of motion for Green's functions
$G_1 \equiv \lan {\color{G}S}^{+(a_1)}| {\color{G}S}^{-(a_1)}\ran,\;\; G_2 \equiv \lan
{\color{BurntOrange}S}^{-(a_2)}|{\color{G}S}^{-(a_1)}\ran$, and employing RPA decoupling procedure
\cite{EPJB,Kuntz,Nolting,Milman,Milman2,Milman3}, we find 
one-mag\-non energies
\be
E(\bm k) = z_1 J_1
\langle S^z \rangle \sqrt{\left[ 1-p\;\frac{z_2}{z_1}(1-\gamma_{2}(\bm{k}))\right]^2 - \gamma_{1}(\bm k)^2},
\label{MagnonNel}
\ee
where
\bea
\hspace*{-0.5cm}\gamma_{1}(\bm k) = \frac{1}{z_1} \sum_{\bm \lambda} \mbox{e}^{i \bm k \cdot
\bm \lambda}, \;\;\; \gamma_{2}(\bm k) = \frac{1}{z_2} \sum_{\bm \delta} \mbox{e}^{i \bm k \cdot
\bm \delta}.  \label{GamaNel}
\eea
In the absence of external magnetic field, the magnon spectrum
is doubly degenerate
as there are two types of magnons with energies given in (\ref{MagnonNel}).
Figure \ref{NEl3D} shows the $k_z = 0$ intersection of    reduced magnon energies 
$\omega (\bm k) = E(\bm k)/(J_1 z_1 \langle S^z \rangle)$.
It can be clearly seen that the spectrum contains the Goldstone mode.
An important feature of RPA decoupling scheme is magnon energy renormalization.
Compared to LSW results \cite{Schmidt}, RPA magnon energies are 
renormalized  by a factor of $\langle S^z \rangle /S$, which includes
the effects of magnon-magnon interactions in RPA. 
As a consequence, the plot of sublattice magnetization from
Figure \ref{MagnFig} suggests that elementary excitations
are not well defined starting at $p \sim 0.7$.

\subsection{AF2 (stripe) phase}

Next we turn to the AF2 phase.
The spin Hamiltonian is
 \bea
\hspace*{-0.8cm}\hat{H}_{\tm{AF2}} & = & J_1 \sum_{\bm n, \bm \lambda_{13}} 
 \hat{{\color{Purple}\bm S}}_{\bm n}^{(a_1)} \cdot  \hat{{\color{Red}\bm S}}_{\bm n +\bm \lambda_{13}}^{(a_3)}
+ J_1 \sum_{\bm n, \bm \lambda_{14}} 
 \hat{{\color{Purple}\bm S}}_{\bm n}^{(a_1)} \cdot  \hat{{\color{G}\bm S}}_{\bm n +\bm \lambda_{14}}^{(a_4)}
\nonumber \\
& + & J_1 \sum_{\bm n, \bm \lambda_{23}} 
 \hat{{\color{BurntOrange}\bm S}}_{\bm n}^{(a_2)} \cdot  \hat{{\color{Red}\bm S}}_{\bm n +\bm \lambda_{23}}^{(a_3)}
+ J_1 \sum_{\bm n, \bm \lambda_{24}} 
 \hat{{\color{BurntOrange}\bm S}}_{\bm n}^{(a_2)} \cdot  \hat{{\color{G}\bm S}}_{\bm n +\bm \lambda_{24}}^{(a_4)}
 \nonumber \\
&+& J_2  \sum_{\bm n, \bm \delta} 
 \hat{{\color{Purple}\bm S}}_{\bm n}^{(a_1)} \cdot  \hat{{\color{BurntOrange}\bm S}}_{\bm n +\bm \delta}^{(a_2)} \;\;\;
+ J_2 \sum_{\bm n, \bm \delta} 
 \hat{{\color{Red}\bm S}}_{\bm n}^{(a_3)} \cdot  \hat{{\color{G}\bm S}}_{\bm n +\bm \delta}^{(a_4)} .
 \label{2.6}
  \eea
\begin{figure}
\bc 
\includegraphics[scale=0.55]{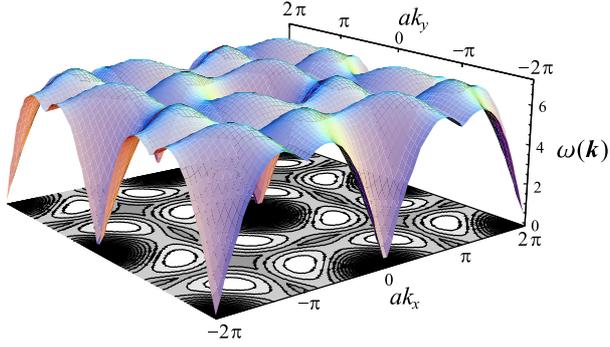}
\caption
{\label{Kol3D} (Color online) 
Reduced magnon energies $\omega_1 (\bm k) = E_1(\bm k)/(J_2 z_2 \langle S^z \rangle)$ 
for $k_z = 0$, $p = 1$ in AF2 phase.}
\ec
\end{figure}
Each of the sublattices in AF2 phase consists of $N_{\tm{S}}=N/4$ sites so that
$\{\bm \lambda_{ij}\}$ connects first neighbours from sublattices $a_i$ and $a_j$ and 
$\{\bm \delta\}$ are   the same vectors as in AF1 phase. Also, it is
seen from Figure \ref{Fig_1} that
all lattice sites in AF2 phase have equal number ($z_1 = 4$)
of  nearest neighbours in two different sublattices.

Using equations of motion for the  Green's functions
$\mathcal{G}_1 \equiv \lan{\color{Purple}S}^{{\color{Purple}+}(a_1)}|
 {\color{Purple}S}^{{\color{Purple}-}(a_1)}\ran,\;\; \mathcal{G}_2 \equiv \lan
{\color{BurntOrange}S}^{{\color{BurntOrange}-}(a_2)}|
{\color{Purple}S}^{{\color{Purple}-}(a_1)}\ran,\;\; \mathcal{G}_3 \equiv \lan
{\color{Red}S}^{{\color{Red}+}(a_3)}|{\color{Purple}S}^{{\color{Purple}-}(a_1)}\ran$,
$\mathcal{G}_4 \equiv \lan {\color{G}S}^{{\color{G}-}(a_4)}|
{\color{Purple}S}^{{\color{Purple}-}(a_1)} \ran$, together with RPA linearization,
 we find renormalized one-magnon energies
\bea
E_{1,2}(\bm k)  =  z_2 J_2
\langle S^z \rangle \sqrt{ \left[1-\gamma_2(\bm k)^2\right] \pm \left(\frac{z_1}{z_2 p}\right)^2 K(\bm k)},  
\label{MagnonKoli}
\eea
where $K(\bm k)$ is the positive square root of
\bea
\left[ \Gamma_1(\bm k)^2-\Gamma_1^*(\bm k)^2 \right]^2 + 4 \left(\frac{ p z_2}{z_1}\right)^2
|\Gamma_1^*(\bm k) - \Gamma_1(\bm k) \gamma_2(\bm k)|^2 .
\eea
Also 
\bea
\Gamma_1(\bm k) & = & \frac{1}{z_1} \sum_{\bm \lambda_{14}}\mbox{e}^{i \bm k \cdot
\bm \lambda_{14}} = \frac{1}{z_1} \sum_{\bm \lambda_{23}}\mbox{e}^{i \bm k \cdot
\bm \lambda_{23}}  \nonumber \\
\Gamma_1^*(\bm k) & = & \frac{1}{z_1} \sum_{\bm \lambda_{13}}\mbox{e}^{i \bm k \cdot
\bm \lambda_{13}} = \frac{1}{z_1} \sum_{\bm \lambda_{24}}\mbox{e}^{i \bm k \cdot
\bm \lambda_{24}}
\eea
while $\gamma_2(\bm k)$ is defined in (\ref{GamaNel}).
Each of magnon energies from (\ref{MagnonKoli}) is doubly degenerate
so that the total number of magnon flavours in the stripe phase is four.
All four magnon branches survive in the limit $J_1 \rightarrow 0$ (i.e. $p \rightarrow \infty$) when 
system described by (\ref{2.6}) reduces to the two decoupled antiferromagnets
with N\' eel order on simple cubic lattices. The magnon spectrum (\ref{MagnonKoli})
then  simplifies to $E(\bm k) = J_2 z_2 \langle S^z \rangle \sqrt{1 - \gamma_2(\bm k)^2}$
so that all four flavours have the same dispersion.
As an illustration of the magnon spectrum in the stripe phase, 
we plot $ \omega (\bm k) = E_1(\bm k)/(J_1 z_2 \langle S^z \rangle)$
for $k_z = 0$ and $p = 1$ on Figure \ref{Kol3D}.
As in the AF1 phase, magnon-magnon interactions induced by RPA
renormalize the magnon energies. It is seen from Figure \ref{MagnFig}
that instability of AF2 phase starts at $p \lesssim 0.7$.

Now that the renormalized  magnon spectrum is determined, 
including effects of frustration,
we can examine its influence
on thermodynamics  of spin $\frac 12$ bcc lattice
 $J_1-J_2$ Heisenberg antiferromagnet.
We have to bear in mind, however, that for correct calculation of thermodynamic
properties in RPA, one must take the full set of the poles of Green's functions
from single system of equations \cite{PhysicaA}. These are $\{ E(\bm k), -E(\bm k)  \}$ in
the N\'{e}el phase  and $\{ E_1(\bm k), -E_1(\bm k),  
 E_2(\bm k),  -E_2(\bm k)  \}$ in the stripe phase. We stress once 
 again that only positive poles of the Green's functions represent
 energies of physical magnons.

\section{Sublattice magnetization} \label{SublMagn}

\begin{figure}
\bc 
\includegraphics[scale=0.50]{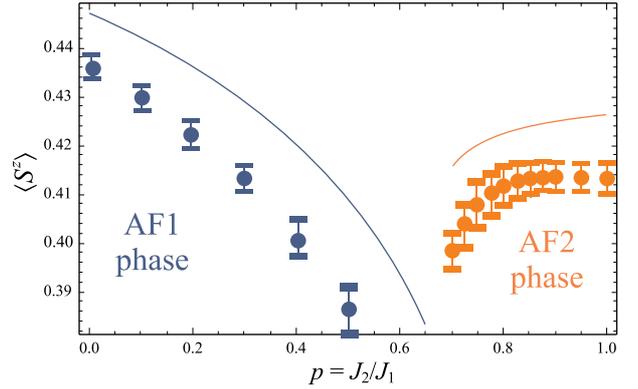}
\caption
{\label{MagnFig} (Color online) 
Sublattice magnetization at $T=0$K in AF1 and AF2 phase 
(Equations (\ref{SigmaRPAN}) and (\ref{SigmaRPAC})). The curves represent RPA result,
while dots are linked-cluster values expansions taken from \cite{Oitma}}
\ec
\end{figure}

Standard spectral theorem enable us to  find the ground state sublattice magnetization
in AF1 phase
\bea 
\langle S^z \rangle_0 & = & \frac 12 \left( \frac{1}{N_{\tm N}} \sum_{\bm
k} \frac{\varepsilon_1(\bm k)}{E(\bm k)}  \right)^{-1}, \label{SigmaRPAN} 
\eea
where
\bea 
\varepsilon_1(\bm k) & = & \langle S^z \rangle \left[z_1 J_1  - z_2 J_2 + z_2 J_2  \gamma_2(\bm k) \right], 
\eea
and in AF2 phase
\bea 
\hspace*{-1.0cm}&&\langle S^z \rangle^{-1}_0 = \frac{p}{N_{\tm S}}\sum_{\bm k}
\frac{1}{K(\bm k)K_1(\bm k) K_2(\bm k)} \label{SigmaRPAC}  \\
\hspace*{-1.0cm}& \times & \left\{ \left[ 6[K(\bm k)+2 |\Gamma_1(\bm k)|^2]  - 
[\Gamma_1(\bm k)^2+ \Gamma_1^*(\bm k)^2] \gamma_2(\bm k) \right]  K_2(\bm k)   \right. \nonumber \\
\hspace*{-1.0cm}& + &  \left. \left[ 6[K(\bm k)-2 |\Gamma_1(\bm k)|^2]  + 
[\Gamma_1(\bm k)^2+ \Gamma_1^*(\bm k)^2] \gamma_2(\bm k) \right]  K_1(\bm k)   \right\}, \nonumber \\
\hspace*{-1.0cm}&&K_{i}(\bm k) = \frac{E_{i}(\bm k)}{z_2 J_2 \langle S^z \rangle }, \hspace{1cm} i = 1,2.
\eea
the plot of which is given at   at Figure \ref{MagnFig}. As noted earlier
in the text, the behaviour of antiferromagnetic order parameter suggest
the phase transition between AF1 and AF2 phase at $p \approx 0.7$.
We also observe that RPA results for sublattice magnetization 
are in agreement with high-order linked-cluster
expansions at $T=0$ of Oitmaa and Zheng \cite{Oitma}, since relative
difference between these methods  is $\approx 2\%$. Similarly, 
our values for the order parameter in AF1 and AF2 phase are 
quite close to the ones obtained from
self-consistent  non-linear spin wave theory \cite{Majumdar1}. For example, it can
be shown \cite{JPA} that for $J_2 =0$, equation (\ref{SigmaRPAN}) gives
$\langle S^z \rangle_0 = \frac 12 \left[_4 F_3 \left(\frac12, \frac12,\frac12,\frac12;1,1,1;1\right) \right]^{-1}
\approx 0.446973$, while non-linear self-consistent spin-wave theory yields
$\langle S^z  \rangle_0 = 1-(1/2) \;_4 F_3 \left(\frac12,\frac12,\frac12,\frac12;1,1,1;1\right)\approx 0.440682$.
Here, $_4 F_3\left(\frac12,\frac12,\frac12,\frac12;1,1,1;1\right)$ denotes the hypergeometric function.

\section{N\' eel Temperature and Phase Diagram} \label{NeelPhaseDiag}

Following \cite{EPJB,Kuntz,Nolting}, we find the   critical temperature in the N\' eel phase 
\bea 
\hspace*{-0.9cm}T_{\tm N}^{\tm{AF1}}(p) & = & \frac{2 J_1}{ I(p)}  \label{TNN} \\
I(p) & = & \frac{1}{N_{\tm N}} \sum_{\bm k}
 \frac{1- p [z_2/z_1][1-\gamma_2(\bm k)]}{\{  1-p[z_2/z_1][1-\gamma_2(\bm k)]  \}^2- \gamma_1(\bm k)^2} \nonumber.
\eea
The general RPA expression for critical temperature remains the same, but in the stripe phase
integral $I(p)$ gets replaced by $\mathcal{I}(p)$, defined by
\bea
\hspace*{-1.1cm}&&\frac{p}{N_{\tm S}} \sum_{\bm k}
\frac{6 [K(\bm k)+2 |\Gamma_1(\bm k)|^2]  - 
[\Gamma_1(\bm k)^2+ \Gamma_1^*(\bm k)^2] \gamma_2(\bm k)}{K(\bm k) K_1(\bm k)^2} \nonumber \\
\hspace*{-1.1cm}& + & \frac{p}{N_{\tm S}} \sum_{\bm k}
\frac{6 [K(\bm k)-2 |\Gamma_1(\bm k)|^2]  + 
[\Gamma_1(\bm k)^2+ \Gamma_1^*(\bm k)^2] \gamma_2(\bm k)}{K(\bm k) K_2(\bm k)^2} \nonumber .
\eea
\begin{figure}
\bc 
\includegraphics[scale=0.50]{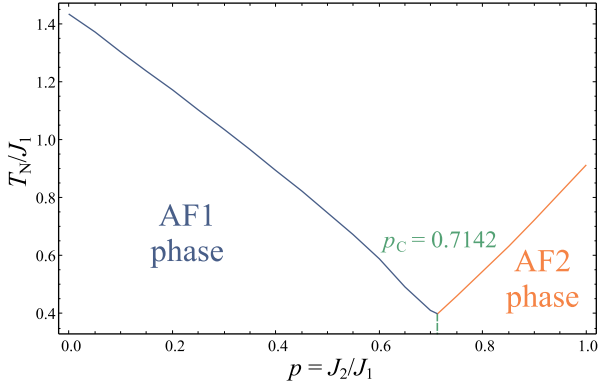}
\caption
{\label{PhaseDiag} (Color online) 
The phase diagram of spin-$\frac 12$ bcc lattice $J_1-J_2$ Heisenberg model}
\ec
\end{figure}
Figure \ref{PhaseDiag} shows our results for reduced critical temperature ($T_{N}/J_1$) in terms 
of frustration ratio $J_2/J_1$. The indication of first-order
transition at $T=0$K and $p_c \approx 0.7$ from Figure \ref{MagnFig}
is further justified by inspection of the phase diagram (see Figure \ref{PhaseDiag}),
giving $p_c =  0.714$.
The blue and orange lines represent AF1-paramagnetic and AF2-paramagnetic
transition, respectively, while the vertical (green) line
shows AF1-AF2 transition line connecting the critical end point 
and the phase transition at $T=0$K.
RPA phase diagram agrees rather well with the one obtain by 
high-temperature series expansion \cite{Oitma}. The relative difference 
between high-temperature series expansion and RPA values
for the critical frustration ratio illustrates this nicely, since
it is  $\approx 1.27 \%$.

\section{Conclusion} \label{Concl}

In summary, we investigated the magnetic properties of
spin $\frac 12$ bcc lattice $J_1-J_2$ Heisenberg antiferromagnet .
Using the method of TGF's, we obtained the phase diagram
as a function of the frustration ratio $p=J_2/J_1$.
Our calculations indicate instability of the long range order
in AF1 and AF2 phase for $p_c= 0.714$. This is in good 
agreement  with previous results found by  high-temperature series expansion \cite{Oitma}.
Also, the RPA predictions for sublattice magnetization
are very close to  high-order linked-cluster
expansions  at $T=0$ \cite{Oitma} and non-linear self-consistent
spin wave theory \cite{Majumdar1}. The main advantage of RPA TGF method
over previously quoted ones is  its successful applicability
on both low and high temperatures.

\subsection*{Acknowledgment} This work was supported by the Serbian
Ministry of Education and Science: Grant No 171009.

\end{document}